\documentclass[aps,prb, twocolumn, superscriptaddress]{revtex4-1}
\usepackage{graphicx}
\usepackage{amsmath}
\usepackage{wasysym}
\usepackage{mathrsfs}
\usepackage{times}
\usepackage{bm}
\usepackage{amssymb}
\usepackage{subfigure}
\usepackage{epstopdf}
\usepackage{hyperref}

\usepackage{enumerate}
\usepackage{color}
\newcommand{\beq}{\begin{equation}}
\newcommand{\eeq}{\end{equation}}
\newcommand{\bea}{\begin{eqnarray}}
\newcommand{\eea}{\end{eqnarray}}

\begin{document}
\draft
\title{Generalizing Quantum Hall Ferromagnetism to Fractional Chern Bands}
\author{Akshay Kumar}
\email{akfour@princeton.edu}
\affiliation{Department of Physics, Princeton
University, Princeton, NJ 08544, USA}
\author{Rahul Roy}
\affiliation{Department of Physics and Astronomy, University of California, Los Angeles, California 90095-1547, USA}
\author{S. L. Sondhi}
\affiliation{Department of Physics, Princeton
University, Princeton, NJ 08544, USA}

\begin{abstract}
We study the interplay between quantum Hall ordering and spontaneous sublattice symmetry breaking in multiple Chern number bands at fractional fillings. Primarily we study fermions with repulsive interactions near half filling in a family of square lattice models with flat C=2 bands and a wide band gap. By perturbing about the particularly transparent limit of two decoupled C=1 bands 
and by exact diagonalization studies of small systems in the more general case, we show that the system generically breaks sublattice symmetry
with a transition temperature $T_c>0$ and additionally exhibits a quantized Hall conductance of  $e^2/h$ as $T \rightarrow 0$. We note the close analogy to quantum Hall ferromagnetism in the multi-component  problem and the connection to topological Mott insulators.  We also discuss generalizations to other fillings and higher Chern numbers.

\end{abstract}
\maketitle

\section{Introduction}

The lowest Landau level plays host to a variety of interesting quantum Hall (QH) phases which are  associated with the  integer and fractional quantum Hall effects.  These arise in 2d electron gases in a high  magnetic field when the electron density is low enough that one or a few of the lowest Landau levels are fully or partially filled. Under these circumstances, details of the periodic potential from the crystal lattice play little role in the physics and can effectively be captured in a single parameter (the effective mass for semiconductor heterostructures or the velocity of the Dirac fermions in graphene) . In multicomponent QH systems at integer fillings, ground states that have topological order and also spontaneously break symmetries can arise - a phenomenon termed QH ferromagnetism (QHFM) ~\cite{sondhi_skyrmions_1993, jungwirth_pseudospin_2000}. An important
feature of such states is that the broken symmetry can persist to nonzero temperatures even as quantum Hall order is lost, as in the celebrated case of bilayer systems at filling factor
$\nu=1$ ~\cite{champagne_evidence_2008}.

Topological bands where the periodic potential does play an important role have been another area of intense recent
interest. These include the $Z_{2}$ invariant bands in topological insulators
(which are associated with time reversal symmetry) which have arguably garnered the most interest ~\cite{hasan_colloquium:_2010}.  Apart from these, topological bands in 2d also include Chern
bands which arise when time reversal symmetry is broken and are close analogs of Landau levels ~\cite{haldane_model_1988}. The analogy is  especially close in the case where the band dispersion
is minimal compared to the other energy scales in the problem. Under these circumstances, fractionally filled Chern bands have been shown to
host precise analogs of Landau level fractional QH phases
~\cite{neupert_fractional_2011,sun_nearly_2011,sheng_fractional_2011,regnault_fractional_2011,qi_generic_2011, parameswaran_fractional_2012, parameswaran_fractional_2013}. While flat Chern bands with $C=1$ share the same global topological invariant, the details of the constituent wavefunctions are considerably different, as for instance reflected in their quantum band geometry ~\cite{parameswaran_fractional_2013}.

In this paper we study analogs of quantum Hall ferromagnets in fractionally filled Chern bands, specifically in Chern bands with Chern number $C >1$. These are states that exhibit topological order at $T=0$ and discrete symmetry breaking for $0 \le T \le T_c$, specifically they break discrete sublattice symmetries. As such they generalize recent theoretical work in the quantum Hall effect
wherein a discrete global symmetry  acts simultaneously on an internal and a spatial degree of freedom ~\cite{abanin_nematic_2010, kumar_microscopic_2013}. This situation occurs in multi-valley systems where different valleys are related by a discrete rotation and gives rise to interesting phenomena. For example in case of AlAs heterostructure, the Hamiltonian has $\mathbb{Z}_2$ symmetry which involves the operation of $\pi/2$ rotation in real space combined with interchange of the 2 valley indices. There is an interesting interplay between the Ising order and topological order. In a clean system, ferromagnetism onsets via a finite temperature Ising transition and exists without topological order at $T>0$. By contrast
disorder induces a random field acting on the Ising order parameter destroying the Ising order but leaving the topological order intact; the resulting phase called the QH random field paramagnet (QHRFP).

In the following we will see that these features do arise in fractionally filled Chern bands as well, specifically in multiple Chern number bands. As has been noted previously ~\cite{bergholtz_topological_2013,qi_generic_2011} and we will
sharpen below, multiple Chern number bands resemble Landau levels with multiple components. A central part of our analysis will be working in a limiting case where this
analogy is sharp. Specifically, we will focus on a family of square lattice models with flat $C=2$ bands and a wide band gap at $1/2$ filling. We show that nearest neighbor density-density repulsive interactions pick QH Ising ferromagnets as ground states (Sec.~\ref{II},~\ref{III}). We also study properties of domain walls germane to the QHRFP phase on lattice (Sec.~\ref{II},~\ref{III}) and discuss the alternative interpretation of the states considered in this paper as topological Mott insulators.
Our ideas can be generalized to flat $C=n>2$ bands at $1/n$ filling (Sec.~\ref{IV}) and also to fillings hosting quantum Hall states with fractional Hall conductance (Sec.~\ref{V}). We close with a summary in Sec.~\ref{VI}.


Before proceeding we would like to draw the reader's attention to two related pieces of work in single Chern ($C=\pm 1$) band systems. Neupert et al ~\cite{neupert_topological_2012,neupert_fractional2_2011} have studied a model for $Z_2$ insulators with an additional global Ising symmetry---but now at fractional fillings and with a Hubbard interaction. They showed that the system exhibits Ising ferromagnetic order along with quantum Hall ordering at fillings $1/2$ and $2/3$. Kourtis and Daghofer ~\cite{kourtis_combined_2013}  have presented numerical results indicating the coexistence of charge density wave order and quantum Hall order at filling $2/5$ of a $C=\pm 1$ band system.

\section{A special flat C=2 band at 1/2 filling} \label{II}



We consider $C=2$ band Hamiltonians on a square lattice. The analogy between
$C=2$ bands and multi-layer/flavor quantum Hall systems is especially transparent at a special set of points in the space of $C=2$ band Hamiltonians.  At these points the Hamiltonian can be written as the tensor sum of Hamiltonians associated with the two inter-penetrating square sub-lattices A and B (say) with no hopping between the sites of A and B. The Hamiltonians associated with each sublattice are identical due to translation symmetry and if they each have Chern number, $C=1$, the lower band of the tensor sum Hamiltonian has $C=2$. An example of such a Hamiltonian with $C=2$ with just second nearest neighbor hoppings is given by
\begin{equation}
\label{eq.1}
H_{o} = \sum_{\vec{k}}
\begin{pmatrix}
c^{\dagger}_{\vec{k} \uparrow} &  c^{\dagger}_{\vec{k} \downarrow}
\end{pmatrix}
\begin{pmatrix}
H_{11}(\vec{k}) &  H_{12}(\vec{k})\\
H^{*}_{12}(\vec{k}) & -H_{11}(\vec{k})
\end{pmatrix}
\begin{pmatrix}
c_{\vec{k} \uparrow} \\
c_{\vec{k} \downarrow}
\end{pmatrix}
\end{equation}
where $H_{11}(\vec{k})=m+\cos(k_x+k_y)+\cos(k_x-k_y)$, $H_{12}(\vec{k})=\sin(k_x+k_y) - i\sin(k_x-k_y)$, $\uparrow$ and $\downarrow$ represent two kinds of orbitals at every lattice site and $m$ is a tunable parameter which leads to a $C=2$ lower band in the range $(-2,0)$. (We take the lattice constant $a=1$.)

$H_{o}$ can be rewritten as $\Delta (1-t+t E_{+,\vec{k}})| \gamma_{+,\vec{k}} \rangle \langle \gamma_{+,\vec{k}} | + (-1+t+t E_{-,\vec{k}})| \gamma_{-,\vec{k}} \rangle \langle \gamma_{-,\vec{k}} |$ where $t=1$, $\Delta=1$, $| \gamma_{-,\vec{k}} \rangle$ and $| \gamma_{+,\vec{k}} \rangle$ are the eigenfunctions for the lower and upper band, respectively and $E_{-,\vec{k}}$, $E_{+,\vec{k}}$ are their corresponding eigenvalues. The eigenstates are related to the old set of states by $|c_{\eta, \vec{k} } \rangle = \sum_{\kappa=-,+} u_{\eta \kappa}(\vec{k}) |\gamma_{\kappa,\vec{k}} \rangle$. If we tune $t \rightarrow 0$ and $\Delta \rightarrow \infty$ (which can always be done by suitably added further nearest neighbor hoppings which preserve the sub-lattice tensor structure) to end up with a flat $C=2$ lower band separated from the upper band by an infinitely large band gap.

\begin{figure}[b]
\centering
(a) \subfigure{\includegraphics[width=\columnwidth]{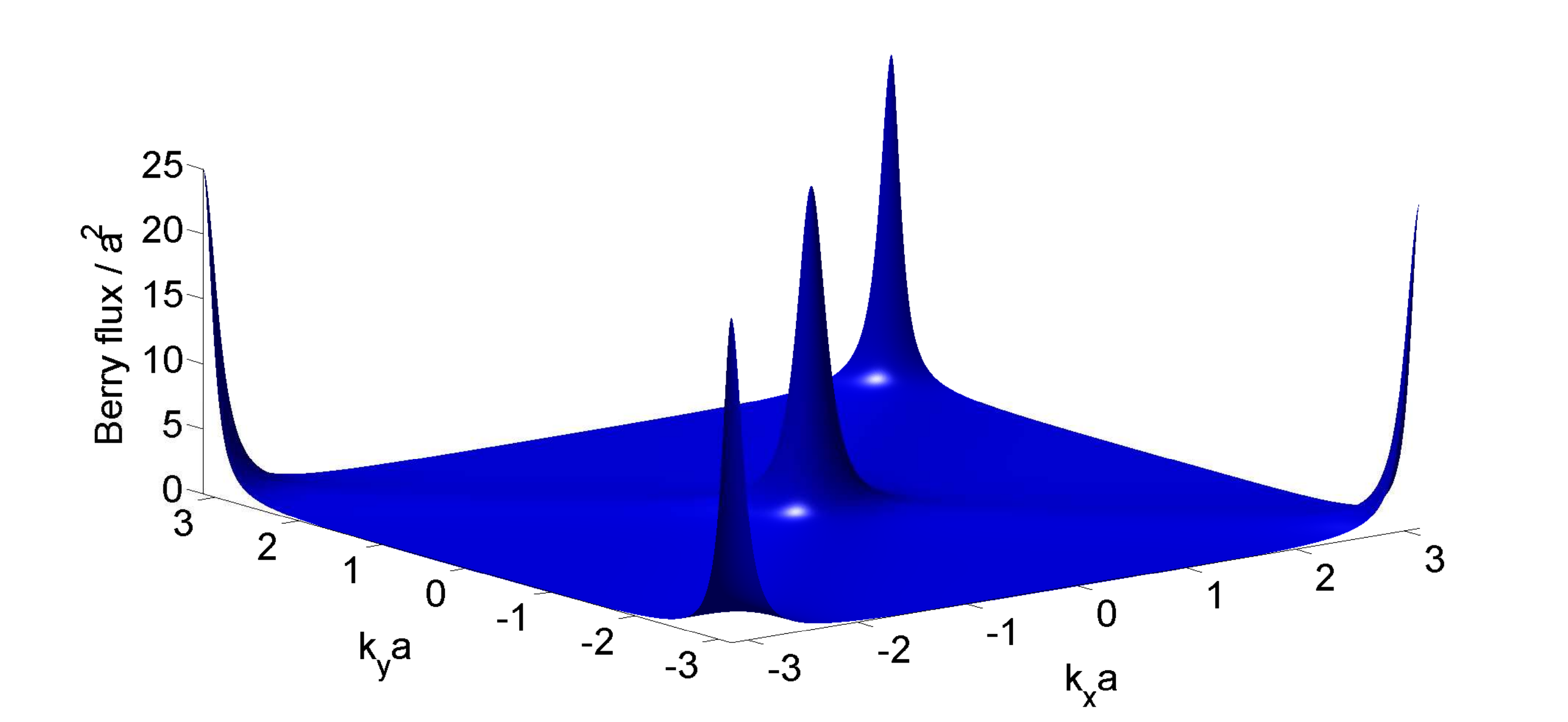}}

(b) \subfigure{\includegraphics[width=\columnwidth]{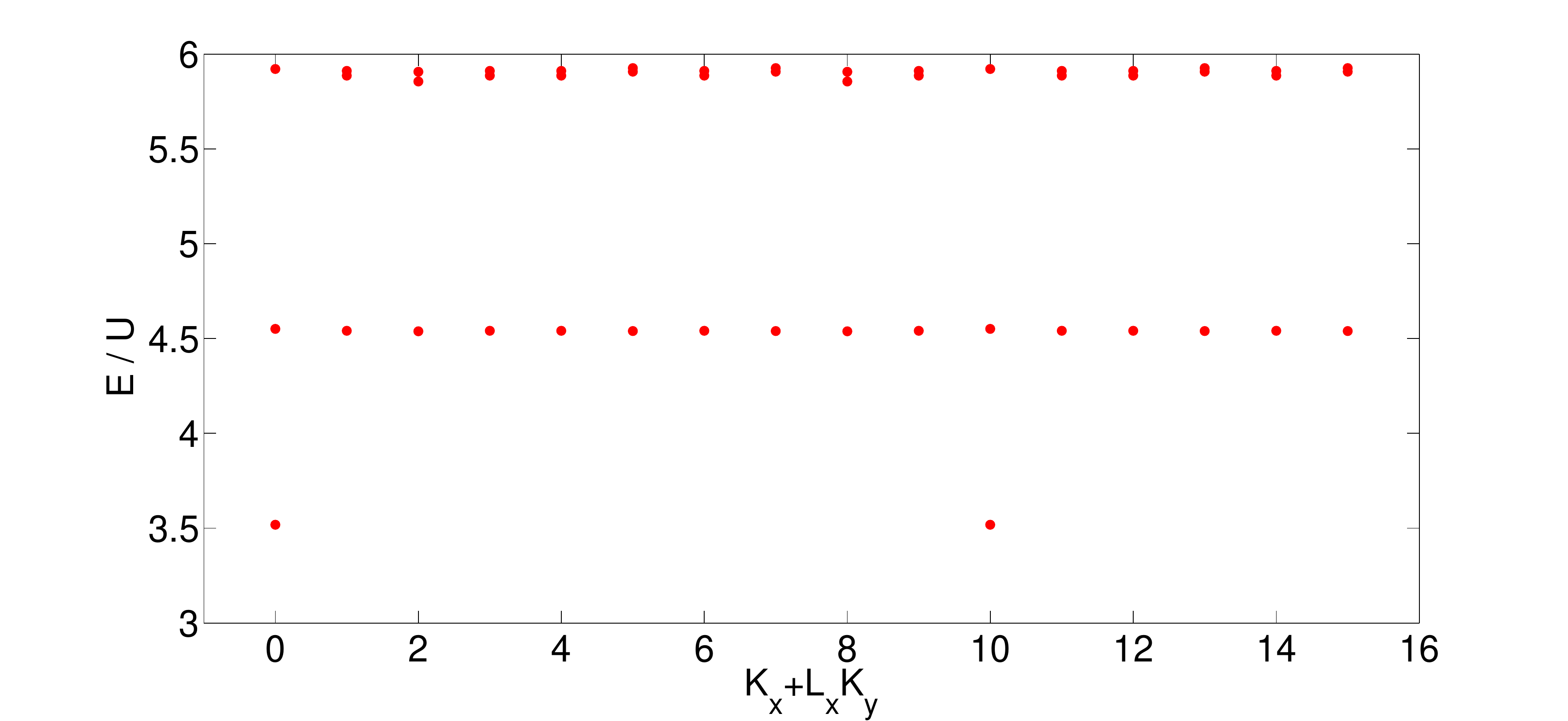}}

\caption{(a) Lower band Chern flux distribution over the Brilliouin zone for the single-particle Hamiltonian $H_{o}$ with $m=-1.8$. (b) Low energy many-body spectrum for 8 fermions on a $4\times4$ lattice for the case of the single-particle part of Hamiltonian chosen as $H_{o}$ with $m=-1.8$ and $V=3U$. (Energies are resolved using total many-body momenta $(K_x,K_y)$ which are in units of $1/a$.) } \label{fig2}
\end{figure}

\begin{figure}[b]
\centering

\subfigure{\includegraphics[width=0.5\columnwidth]{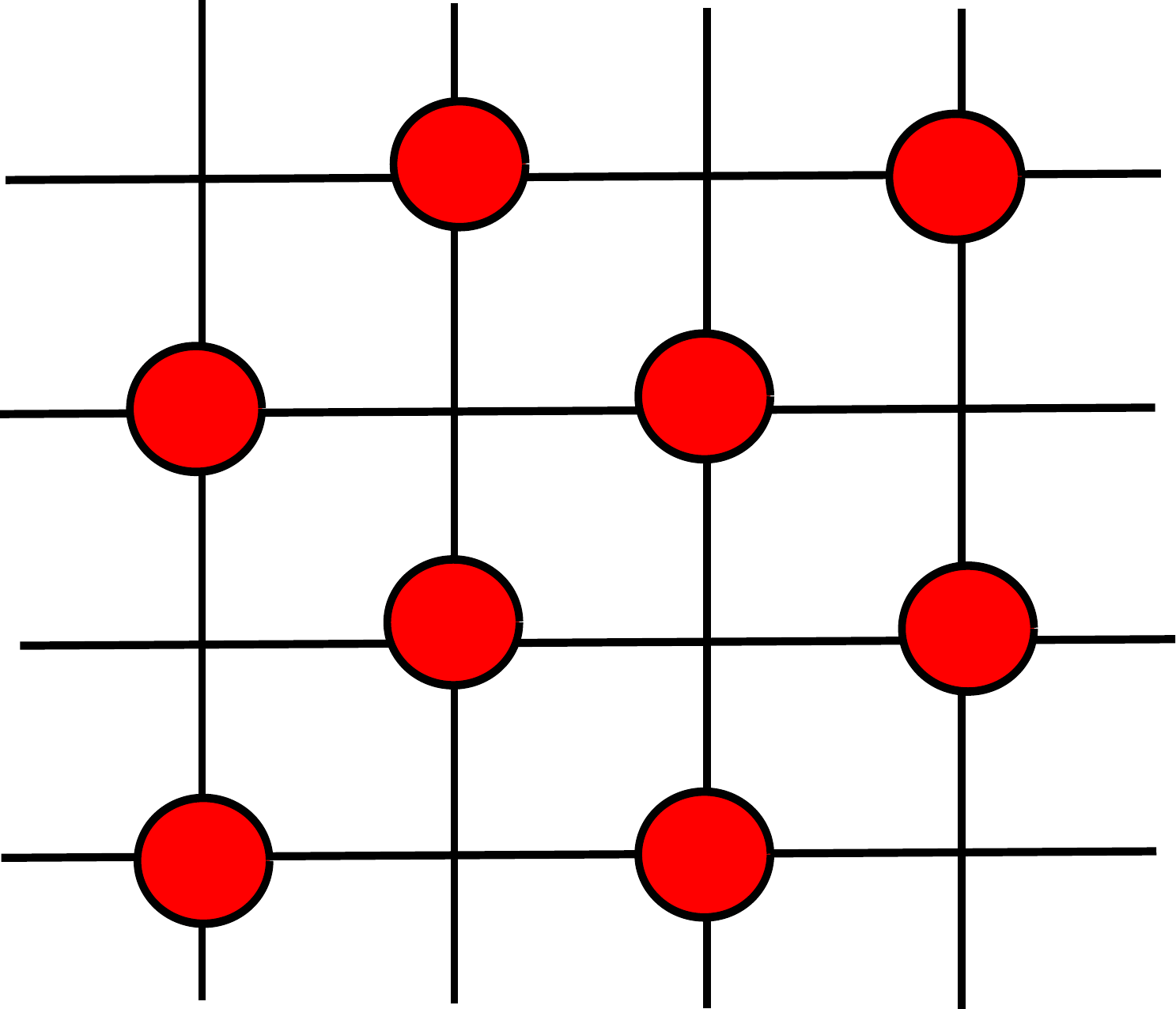}}

\caption{Ising ordered ground state for $H_{proj.}$ at half filling.} \label{fig1}
\end{figure}

We then add interactions:
\begin{equation}
\label{eq.2}
H_{int.} = V \sum_{\vec{i}} n_{\vec{i}\uparrow}n_{\vec{i}\downarrow} + U \sum_{\langle \vec{i} \vec{j} \rangle} \sum_{\kappa,\kappa'=\uparrow,\downarrow} n_{\vec{i}\kappa}n_{\vec{i}\kappa'}
\end{equation}
where $V$ is the on-site,intra-sublattice, interaction strength and $U$ is the nearest neighbor (NN), inter-sublattice,  interaction strength.  Since we are interested only in the
low energy part of the many-body spectrum, we will restrict our analysis to  $H_{int.}$ projected onto the lower flat band. $H_{proj.}$ can be obtained by making the change
$c_{\eta,\vec{k}} \rightarrow u_{\eta -}(\vec{k}) \gamma_{-,\vec{k}}$ in $H_{int.}$. We will work at half filling.

Intuitively we expect two regimes for the Hamiltonian (\ref{eq.2}). At $V \ll U$ the particles can efficiently avoid the NN repulsion by segregating on one sublattice while for
$V \gg U$ they will minimize the on-site repulsion by inhabiting both sublattices. In the first regime, the many-body spectrum has 2 degenerate ground states (Fig.~\ref{fig2}(b)) corresponding to
fully filling the $C=1$ bands that live on sublattice A/B. The reader can check that these states are always eigenstates of the projected $H_{int}$ as they are tensor products
of the empty state on A/B with the fully filled state on B/A. By explicit computation for the case of $m=-1.8$ which we will use for illustration purposes in this paper, we find that they are the ground states for $V < 4.3U$ which covers the physically interesting
regime where $V$ is not smaller than $U$. Both ground states break the Ising symmetry
between the two sublattices or alternatively are $(\pi,\pi)$ charge density waves, one having all its weight on the A lattice and the other on B lattice (Fig.~\ref{fig1}). These results can be interpreted using approximate Slater determinant ground states constructed from single particle wavefunctions which are formed from linear combinations of the Bloch states at a given $\mathbf{k}$  and $\mathbf{k} + (\pi, \pi)$  such that the wavefunctions have support only on one of the lattices. Clearly
both states exhibit a Hall conductance $\sigma_H= {e^2 \over h}$. For $V > 4.3U$ the many-body spectrum does not have two degenerate ground states and hence a phase transition has ocurred.

The excitations about these symmetry broken ground states are two-fold. First there are particle hole excitations in which a hole is created on sublattice A (say) and a particle
is created on sublattice B. While the quasiparticle band is flat, the quasihole band has finite dispersion. The minimum particle-hole pair creation energy asymptotes to a value which depends on the choice
of $C=1$ Hamiltonian. For the case of $m=-1.8$ it asymptotes to $4U-0.6V$. Note that this gap does not close for $V<4.3U$ so the transition that we report in the ground state is first order.

\begin{figure}[b]
\centering
(a) \subfigure{\includegraphics[width=0.5\columnwidth]{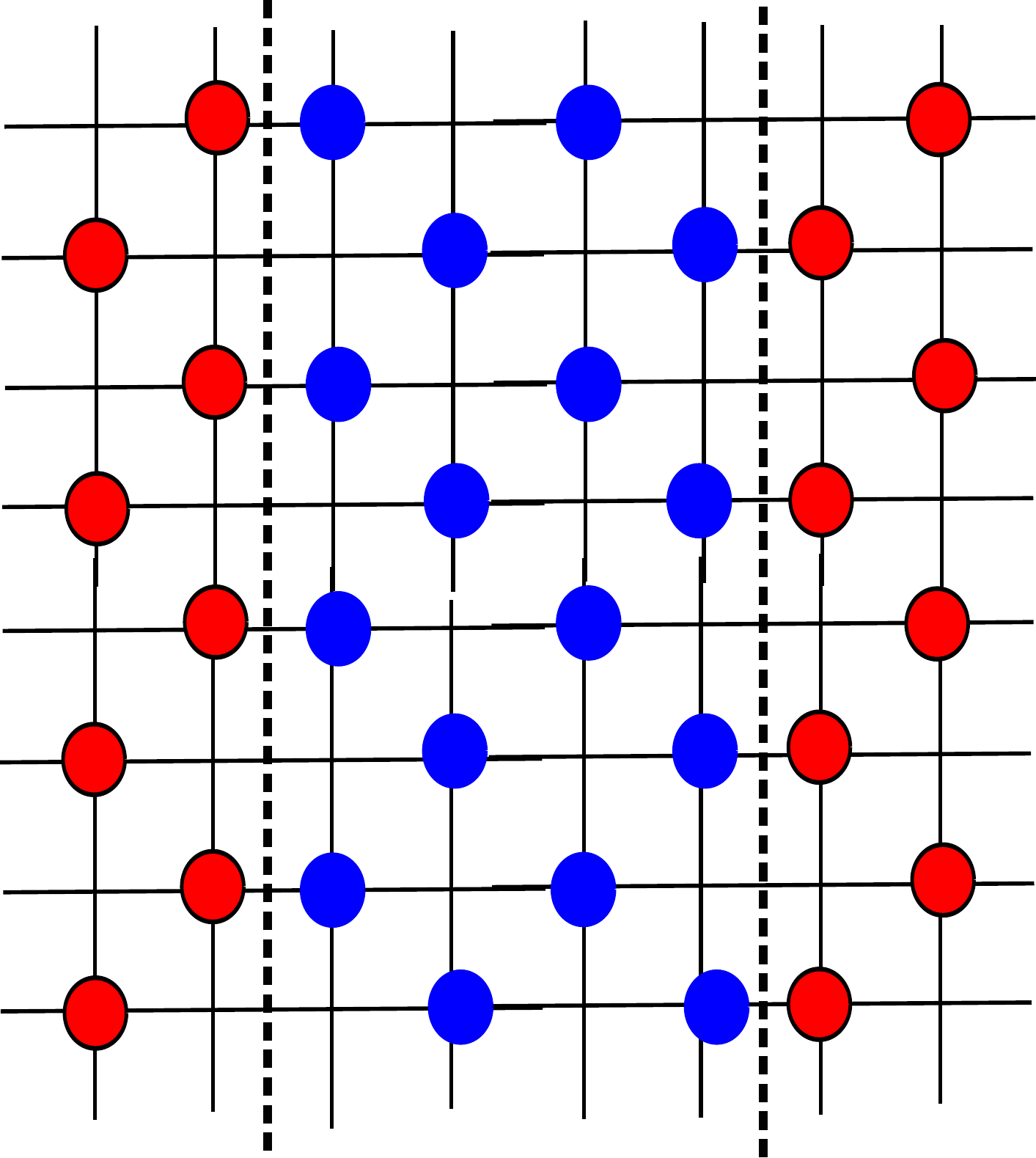}}

(b) \subfigure{\includegraphics[width=0.5\columnwidth]{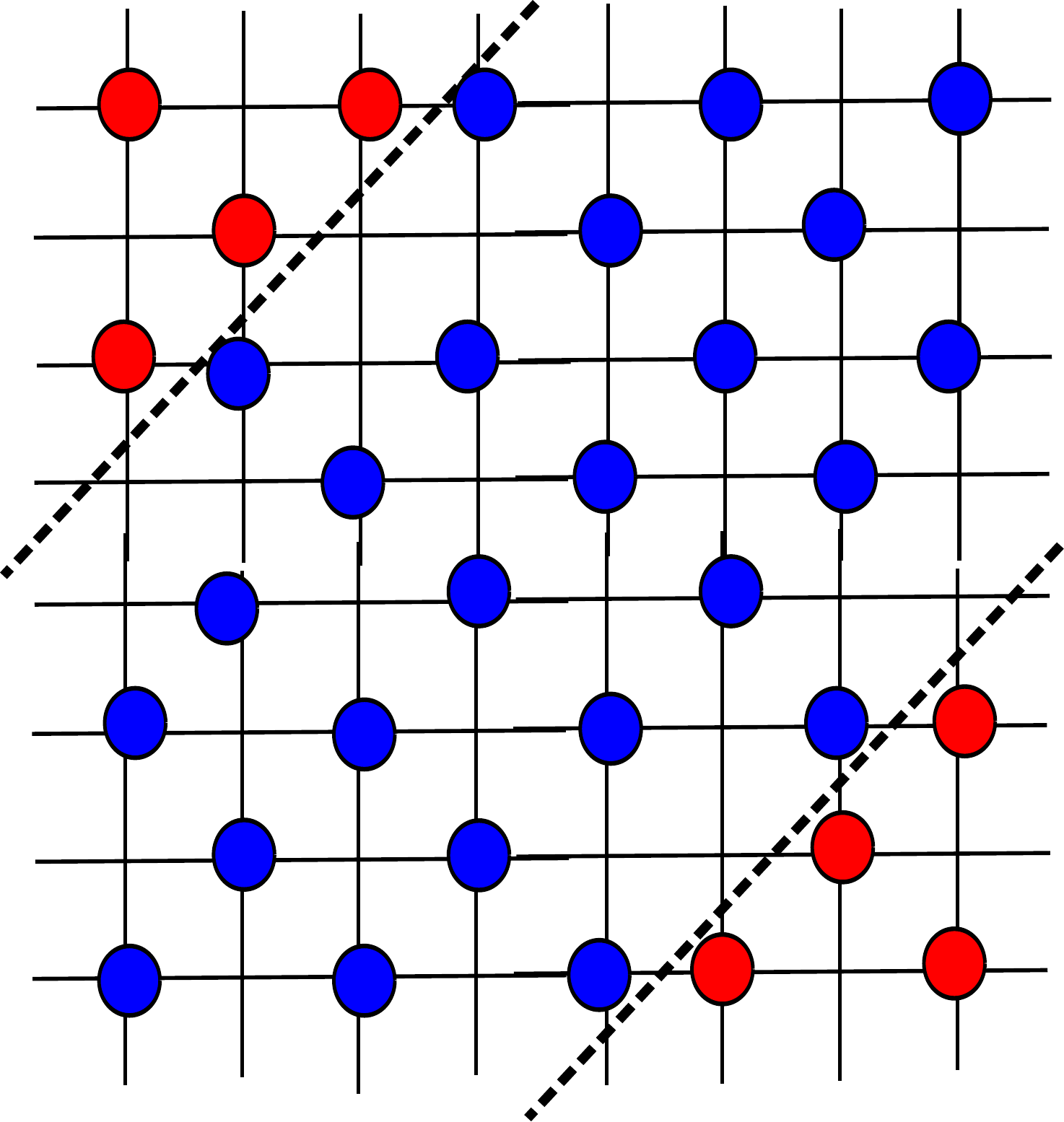}}
\caption{2 species of domain wall considered in the main text.} \label{fig3}
\end{figure}

The second set of excitations are domain walls---across which sub-lattice occupation changes---are the topological defects of the Ising order ferromagnet discussed above. Two of the many possible orientations of these domain walls are depicted in Fig.~\ref{fig3}. The energy of the domain walls is set by the interactions. For example in the case of domain wall oriented as shown in Fig.~\ref{fig3}(b), the energy per unit length is approximately $0.2V+2.4U$ for $m=-1.8$. The electronic structure of these domain walls is
of interest as it reflects the interaction between the topological order and symmetry breaking. In the current limit a pair of counter-propagating gapless, chiral modes of opposite sub-lattice index exist at a domain wall. In the absence of inter-sublattice hopping both sides ``see'' the domain wall as the boundary to a topologically trivial vacuum.

With these excitations identified we can describe the response of the state to temperature, doping and disorder. At half filling in the clean system, topological order is lost at any $T>0$ by the proliferation of particles and holes which, following the standard lore in the QHE, one can think of these as vortices in the topological gauge field. [Admittedly the topological order is somewhat trivial here being that of the Integer Hall effect. We will present a fractional case later.] Domain walls are bounded in size at small $T$ but will proliferate above a critical temperature $T_c$ leading to a finite temperature Ising phase transition. In our present limit domain walls conduct whence we expect the AC longitudinal conductivity to be sizeable on the scale of the domains and to increase sharply in the DC limit around $T_c$. Weak doping around half filling will be accommodated by the
inclusion of a density of particles/holes in the ground state. Absent disorder and for our short ranged model, doping will destroy QH order
while continuing to preserve the sublattice order---whence the finite $T$ physics described above will survive. Finally with disorder two
new effects will enter. First, the physics of the random field Ising model ~\cite{imry_random-field_1975, binder_random-field_1983} will enter
and restore sublattice symmetry at and near half filling. Second, disorder will localize the particles/holes leading to a quantized
Hall effect for a finite range of doping.

We have described the parent state at half filling as a quantum Hall ferromagnet in the above. Here we would like to note that it is also
a species of Mott insulator by which we mean a state with the constituent particles localized on account of strong interactions. Indeed,
for $|m| > 2$, the lower band is topologically trivial ($C=0$) and if we again flatten the bands as described above and add the same  interactions, then the resulting state is reasonably described as a Mott insulator with additional sublattice symmetry breaking. In this
case one can also construct localized Wannier states with support entirely on one of the sub-lattices, A or B and the filled band corresponds to a occupation of all of the Wannier states on one or the other sub-lattice. For our $C=2$ system, Wannier states exponentially localized in both dimensions can no longer be defined. Nevertheless, the resulting ground state is insulating in the bulk and may be regarded as a form of Mott Chern insulator with sublattice symmetry breaking. Topological Mott phases have been previously been proposed in  some very different settings ~\cite{rachel_topological_2010,raghu_topological_2008,hsu_topological_2011}.


\section{Other flat C=2 bands at 1/2 filling} \label{III}

\begin{figure}[b]
\centering
(a) \subfigure{\includegraphics[width=\columnwidth]{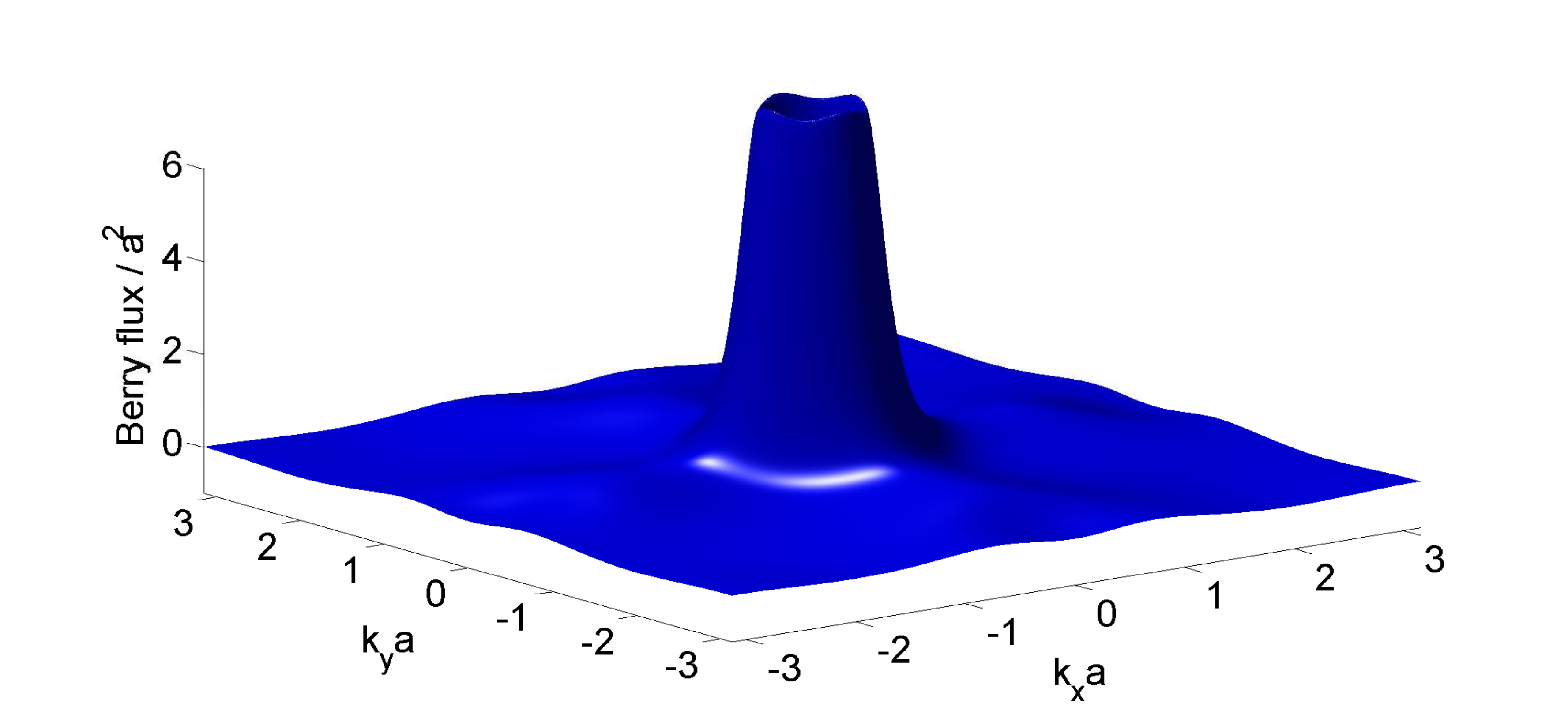}}

(b) \subfigure{\includegraphics[width=\columnwidth]{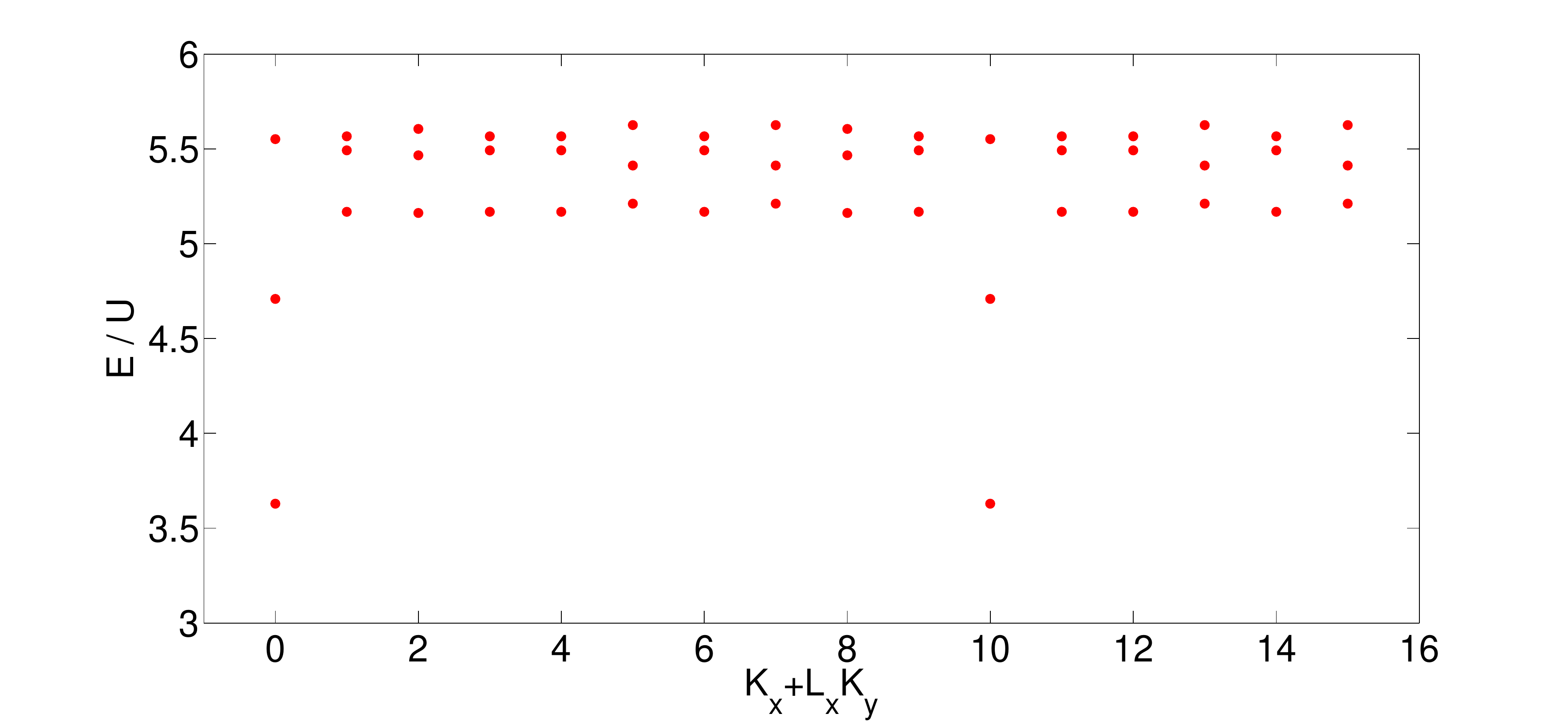}}

\caption{(a) Lower band Chern flux distribution over the Brilliouin zone for the single-particle Hamiltonian $H^{'}_{o}$ with $m=-1.8$. (b) Low energy many-body spectrum for 8 fermions on a $4\times4$ lattice for the case of the single-particle part of Hamiltonian chosen as $H^{'}_{o}$ with $m=-1.8$ and $V=3U$. (Energies are resolved using total many-body momenta $(K_x,K_y)$ which are in units of $1/a$.)} \label{fig4}
\end{figure}

When we move away from the fixed points where the band Hamiltonians sum structure by adding inter-sublattice hopping, we expect the many body gap to generically  remain stable. In Sec.~\ref{II}, the single-particle Hamiltonian is such that the Berry curvature corresponding to the lowest band is symmetric under translations by $(\pi, \pi)$ in the Brillouin zone and is concentrated in two pockets in the Brillouin zone (Fig.~\ref{fig2}(a)). Our earlier statements regarding the ground states should hold irrespective of the kind of Chern flux distribution. As a check, we consider a single-particle Hamiltonian $H^{'}_{o}$ for which the lower band's Chern flux is concentrated in only one pocket (Fig.~\ref{fig4}(a)). (There is always some contribution coming from the other regions of the Brillouin zone.)  This can be made possible by turning on hopping between A and B lattice sites. One such instance can be given in the form of Equation~\ref{eq.1} where $H^{'}_{11}(\vec{k})=m+\cos(k_x)+\cos(k_y)$, $H^{'}_{12}(\vec{k})= \frac{1}{2} (\cos(2k_y)-\cos(2k_x)) + i (\cos(k_x-k_y) - \cos(k_x+k_y))$ and $-2<m<0$.

The new projected Hamiltonian $H^{'}_{proj.}$ is diagonalized for 8 particles on a $4 \times 4$ lattice (Fig.~\ref{fig4}(b)). A 2-fold degenerate ground state and a gap of order $U$ are again observed. We also check for broken sublattice symmetry by computing the Fourier transform of $\langle \rho'(\vec{x}) \rho'(0) \rangle$ where $\rho'$ is the density operator projected onto the lowest flat band. Peaks are observed at $(0,0)$ and $(\pi,\pi)$ indicating the presence of sublattice symmetry breaking (CDW order).

\begin{figure}[b]
\centering
(a) \subfigure{\includegraphics[width=\columnwidth]{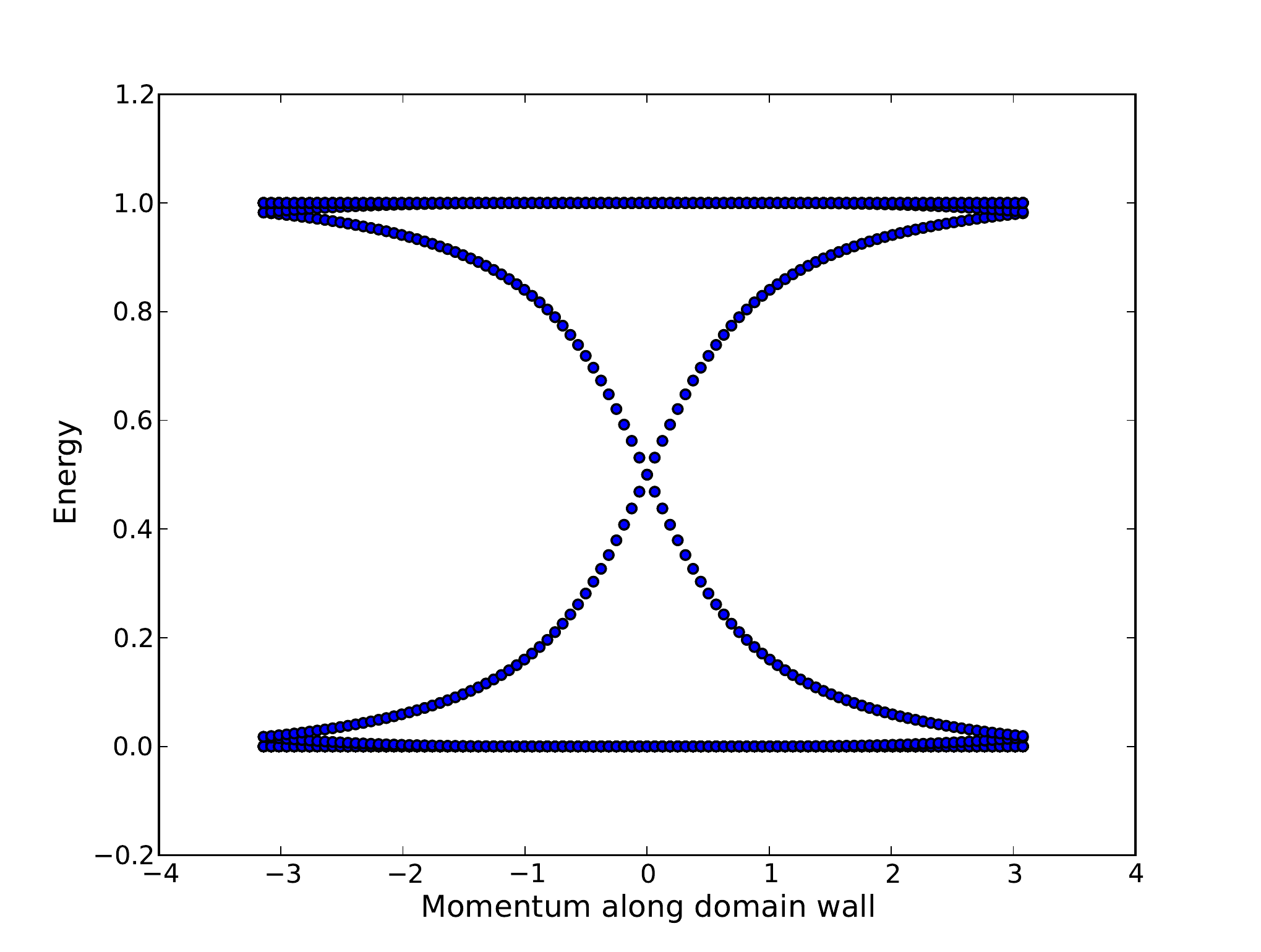}}

(b) \subfigure{\includegraphics[width=\columnwidth]{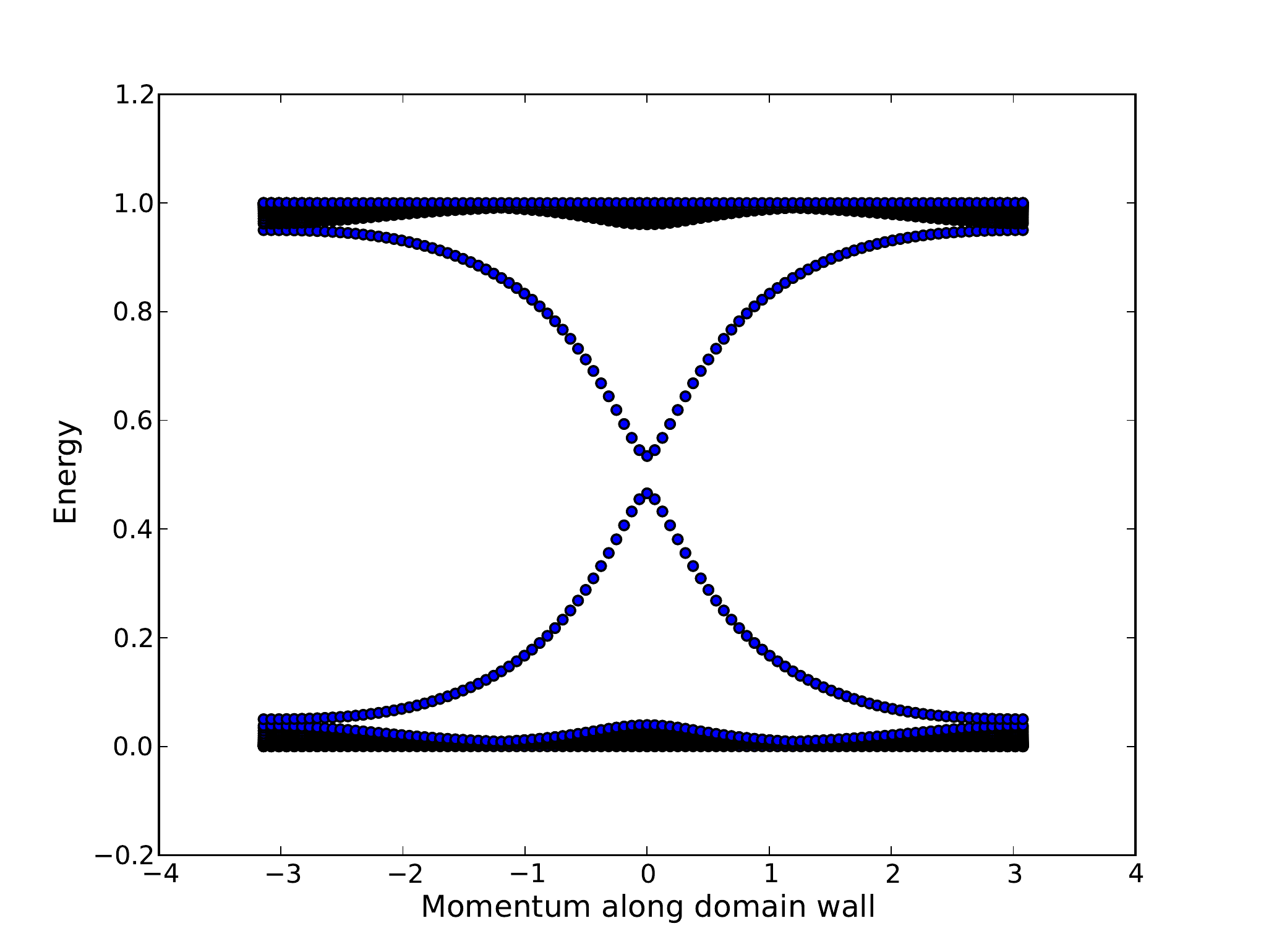}}

\caption{(a) Spectra of the Hamiltonian in Equation~\ref{eq3} with $m=-1$ and choice of orientation of domain wall as shown in Fig.~\ref{fig3}(b). (a) $t = 0$. (b)  $t = 0.3$. (Momentum along domain wall is in units of $1/\sqrt{2}a$.) Similar results are obtained for the other choice of $V(\vec{i})$.} \label{fig5}
\end{figure}																																				 

The response of the state to $T$, doping and disorder is qualitatively the same as before. The one significant difference is in the electronic structure of the domain walls. Upon inclusion of inter-sublattice hopping we may wonder whether they gap generically or are remain protected by some symmetry of the problem. This examination is partly motivated by analogous objects in the Ising quantum Hall FM in the AlAs problem ~\cite{abanin_nematic_2010, kumar_microscopic_2013}
where the domain walls {\it are} gapless to an excellent approximation. To this end we have examined domain walls in the (1,0) and and (1,1)
orientations produced by the added one-body potential
\begin{equation}
\label{eq3}
H=\sum_{\vec{i}} V(\vec{i}) (\rho_{\uparrow}(\vec{i})+\rho_{\downarrow}(\vec{i}))_{projected}
\end{equation}
where the projection is on to the lowest band of a generic Hamiltonian of the form given by Equation~\ref{eq.1}. For purposes of explicit computation below we will take $H_{11}(\vec{k})=m+\cos(k_x+k_y)+\cos(k_x-k_y)$, $H_{12}(\vec{k})=\sin(k_x+k_y) - i\sin(k_x-k_y) - t(\sin(qx)+ i\sin(qy))$ where $t$ is the parameter for hopping between the two sublattices. The $V(\vec{i})$ corresponding to two orientations of domain wall are:
\begin{equation}
\label{eq4}
V_1(\vec{i}) =
\begin{cases}
(-1)^{i_x+i_y}, & i_x<N/4 \\
(-1)^{i_x+i_y+1}, & N/4<i_x<3N/4 \\
(-1)^{i_x+i_y}, & 3N/4<i_x \\
\end{cases}
\end{equation}
\begin{equation}
\label{eq5}
V_2(\vec{i}) =
\begin{cases}
(1+(-1)^{i_x-i_y})/2, & (i_x-i_y)<N/4 \\
(1+(-1)^{i_x-i_y+1})/2, & N/4<(i_x-i_y)<3N/4 \\
(1+(-1)^{i_x-i_y})/2, & 3N/4<(i_x-i_y) \\
\end{cases}
\end{equation}
where $N$ is the number of sites along $x/y$ direction and we are assuming periodic boundary conditions along both $x$ and $y$ directions. In Equation~\ref{eq4}, domain walls run parallel to $y$ direction and in Equation \ref{eq5} they run parallel to the diagonal. In both cases, we find a gap in the spectrum (Fig.~\ref{fig5}) and this should hold for generic orientations of the walls.

\section{Generalization to higher Chern bands} \label{IV}

Our results can be generalized to flat $C=n >2$ bands at filling factor $1/n$ . An obvious way to begin is to make a flat $C=n >1$ lower band by putting $n$ decoupled lattices together, each independently having a flat $C=1$ lower band. Then one can arrange for repulsive interactions to pick the $n$ fully occupied single sublattice bands as ground states. For example in the $C=4$ case, on-site, nearest neighbor, next-nearest neighbor and next-next-neighbor repulsive interactions pick 4 degenerate ground states having both many-body Chern number $C=1$ and broken sublattice/translational symmetry. Such a state is the Chern band analog of a QHFM in a system with a $\mathbb{Z}_n$ global symmetry, examples with $n=3$ are the Si (111) QH system at filling factors $1$ and $5$. In addition to domain walls the system will now exhibit proto-vortices where $n$ distinct domains come together at a point. For $n>4$ the system will also exhibit a $T>0$ Kosterlitz-Thouless phase ~\cite{jose_renormalization_1977}.

\section{Fractional States} \label{V}

Let us return to the $C=2$ band but now turn to filling $1/6$ with the Hamiltonian as the one in Equation~\ref{eq.2} plus a set of further intra-lattice and inter-lattice further neighbor repulsive interaction terms. In the decoupled limit, we can tune the intra-lattice interactions to be those that stabilize a state that  forms a fractional Chern insulator state at $1/3$ of a single $C=1$
band ~\cite{regnault_fractional_2011}. Adding a sufficiently large U and a number of further neighbor inter-lattice repulsive terms of adequate magnitude will ensure that domain walls between regions which reside on one sub-lattice and the other are energetically unfavorable and thus favor a state that breaks sublattice symmetry and exhibits the
topological order of the $\nu=1/3$ Laughlin state and a quantized Hall conductance
$\sigma_H = e^2/(3 h)$. Evidently this construction can be repeated for other known QH states
in a $C=1$ Chern band.


\section{Concluding Remarks} \label{VI}

To summarize, we found analogs of QH ferromagnets in the menagerie of fractional Chern insulator phases. The case of a flat $C=2$ band at $1/2$ filling is analogous to AlAs QH system at filling factor $1$. The two ground states are the broken sublattice symmetry states having topological order. Unlike in the AlAs system the domain walls come naturally with gapped electronic excitations.

In future work it would be interesting to locate the phases that we have discussed in a larger phase
diagram in which we restore dispersion to the band and also allow its Chern number to go through
a transition. This will introduce the trivial Mott insulator/FM and the trivial and topological half
filled metal into the phase diagram and it would be interesting to see exactly how the various phases
fit together and the nature of the transitions between them.


\acknowledgements{We would like to thank Sonika Johri for helpful discussions regarding exact diagonalization calculations and Titus Neupert and Fenner Harper for comments on the manuscript. This work was supported by NSF Grant Numbers DMR 1006608, 1311781 and PHY-1005429 (AK and SLS) and the Alfred P. Sloan Foundation (RR).}


\begin{thebibliography}{99}
\vspace{0mm}

\providecommand{\natexlab}[1]{#1}
\providecommand{\url}[1]{\texttt{#1}}
\expandafter\ifx\csname urlstyle\endcsname\relax
  \providecommand{\doi}[1]{doi: #1}\else
  \providecommand{\doi}{doi: \begingroup \urlstyle{rm}\Url}\fi

\bibitem{sondhi_skyrmions_1993}
S. L. Sondhi, A. Karlhede, S. A. Kivelson, and E. H. Rezayi, Phys. Rev. B. \textbf{47}, 16419 (1993).

\bibitem{jungwirth_pseudospin_2000}
T. Jungwirth, and A.H. MacDonald, Phys. Rev. B. \textbf{63}, 035305 (2000).

\bibitem{champagne_evidence_2008}
A. R. Champagne, J. P. Eisenstein, L. N. Pfeiffer, and West, K. W., Phys. Rev. Lett. \textbf{100}, 096801 (2008).

\bibitem{hasan_colloquium:_2010}
M. Z. Hasan, and C. L. Kane, Rev. Mod. Phys.\textbf{82}, 3045 (2010).

\bibitem{haldane_model_1988}
F. D. M. Haldane, Phys. Rev. Lett. \textbf{61}, 2015 (1988).

\bibitem{neupert_fractional_2011}
T. Neupert, L. Santos, C. Chamon, and C. Mudry, Phys. Rev. Lett. \textbf{106}, 236804 (2011).

\bibitem{sun_nearly_2011}
K. Sun, Z. Gu, H. Katsura and S. Das Sarma, Phys. Rev. Lett. \textbf{106}, 236803 (2011).

\bibitem{sheng_fractional_2011}
D. N. Sheng, Z .C. Gu, K. Sun, and L. Sheng, Nat. Commun. \textbf{2}, 389 (2011).

\bibitem{regnault_fractional_2011}
N. Regnault and B. A. Bernevig, Phys. Rev. X. \textbf{1}, 021014 (2011).

\bibitem{qi_generic_2011}
Xiao-Liang Qi, Phys. Rev. Lett. \textbf{107}, 126803 (2011).

\bibitem{parameswaran_fractional_2012}
S. A. Parameswaran, R. Roy, and S.L. Sondhi, Phys. Rev. B. \textbf{85}, 241308 (2012).

\bibitem{parameswaran_fractional_2013}
S. A. Parameswaran, R. Roy, and S. L. Sondhi, Comptes Rendus Physique \textbf{14}, 816 (2013).

\bibitem{abanin_nematic_2010}
D. A. Abanin, S. A. Parameswaran, and S. A. Kivelson, and S. L. Sondhi, Phys. Rev. B. \textbf{82}, 035428 (2010).

\bibitem{kumar_microscopic_2013}
A. Kumar, S. A. Parameswaran, and S. L. Sondhi, Phys. Rev. B. \textbf{88}, 045133 (2013).

\bibitem{bergholtz_topological_2013}
E. J. Bergholtz and Zhao Liu, Int. J. Mod. Phys. B \textbf{27}, 1330017 (2013).

\bibitem{neupert_topological_2012}
T. Neupert, L. Santos, S. Ryu, C. Chamon, and C. Mudry, Phys. Rev. Lett. \textbf{108}, 046806 (2012).

\bibitem{neupert_fractional2_2011}
T. Neupert, L. Santos, S. Ryu, C. Chamon, and C. Mudry, Phys. Rev. B. \textbf{84}, 165107 (2011).

\bibitem{kourtis_combined_2013}
S. Kourtis and M. Daghofer, arXiv:1305.6948

\bibitem{binder_random-field_1983}
K. Binder, Z. Physik B - Condensed Matter \textbf{50}, 343 (1983).

\bibitem{imry_random-field_1975}
Y. Imry and Shang-keng Ma, Phys. Rev. Lett. \textbf{35}, 1399 (1975).

\bibitem{hsu_topological_2011}
C.H. Hsu, S. Raghu, and S. Chakravarty, Phys. Rev.B. \textbf{84}, 155111 (2011).

\bibitem{raghu_topological_2008}
S. Raghu, Xiao-Liang Qi, C. Honerkamp and Shou-Cheng Zhang, Phys. Rev. Lett. \textbf{100}, 156401 (2008).

\bibitem{rachel_topological_2010}
S. Rachel and K. Le Hur, Phys. Rev. B. \textbf{82}, 075106 (2010).

\bibitem{jose_renormalization_1977}
J. V. Jose, L. P. Kadanoff, S. Kirkpatrick, and D. R. Nelson, Phys. Rev. B. \textbf{16}, 1217 (1977).

\end{thebibliography}
\end{document}